\documentstyle[12pt,twoside,psfig]{article}
\begin{document}
\title{Sonoluminescing Gas Bubbles}
\author{$\ $ \\
{\bf I. Scott$^1$, H.-Th. Elze$^{1,2}$, T. Kodama$^2$ and J.
Rafelski$^1$}\\
$\ $\\
$^1$Department of Physics, University of Arizona, Tucson, AZ
85721\\
\rule{0cm}{0.7cm} and \\ \rule{0cm}{0.6cm} 
$^2$Universidade Federal do Rio de Janeiro, Instituto de
F\'{\i}sica\\
Caixa Postal 68.528, 21945-970 Rio de Janeiro, RJ, Brazil $\ $}
\date{Revised August 1998}
\maketitle
\thispagestyle{empty} \pagestyle{myheadings}
\markboth{Scott, Elze,
Kodama and  Rafelski}{Sonoluminescing Gas Bubbles}
\begin{abstract}
{\noindent  
We draw attention to the fact that the popular but unproven 
hypothesis of shock-driven sonoluminescence is incompatible 
with the reported synchronicity of the single bubble
sonoluminescence (SBSL) phenomenon. Moreover, it is not a 
necessary requirement, since we show that the sub-shock 
dynamic heating in gas bubble cavitation can lead
to conditions required to generate intense 100ps light
pulses. To wit we study the dynamics 
of the interior of a cavitating gas bubble subject to 
conditions suitable for sonoluminescence. We explore  
quantitatively the transfer of energy  from the sound 
wave to the bubble interior, the frequency of atomic
collisions in the bubble, the limits of quasi-stability 
of the non-linear bubble oscillations driven by
an acoustical field, and obtain the implied reaction
time scales.
}
\end{abstract} \hbadness=10000 
PACS: 47.40.Dc, 47.10+g, 43.30.+m, 78.60.Mq
\subsubsection*{1. Introduction}
Current interest in the gas bubble cavitation dynamics has been
revived by the development of experimental conditions
 which allow a single bubble \cite{Gai90}, trapped at the 
anti-node of a standing acoustic pressure wave, 
to emit relatively intense short light pulses \cite{Bar91}.
This single bubble sonoluminescence (SBSL) displays 
considerable sensitivity to the experimental conditions 
and parameters.  For a  summary of the many ensuing 
experimental results we refer to the recent review by 
Barber {\it et.\,al.}\/ \cite{Bar97}. 
In particular they have found that the light
flashes occur within 0.5\,ns of maximum compression, 
last often less than 50\,ps, comprise up to  a million 
photons, and occur with synchronicity exceeding the 
frequency stability of the driving acoustical wave.  
The interplay of the diffusion 
processes and dynamical motion of the bubble has been also 
explored in depth \cite{Brenner} suggesting that
there is ongoing diffusion of the gas between the bubble and the
liquid, and that the gas contents of the SBSL bubble is result of
self-fine-tuning dynamics (rectified diffusion).

The dynamics of the interior of the gas bubble and the energy inflow
from  the surrounding  fluid into the bubble need to be understood
in order to describe the energy focusing process in SBSL which
is capable to convert acoustical energy into light.
For sub-shock, spherical bubble dynamics we develop here  a 
modification of the original Rayleigh-Plesset (RP) cavitation  model
\cite{Nep80}, which addressed only the  dynamics of the spherical 
gas-liquid  interface.  The bubble interior
enters the RP-model in terms of the (average) bubble gas pressure
contributing to the surface dynamics and relevant in particular in the 
collapse phase. Our extension, based  on variational 
approximation of hydrodynamics \cite{YM79,Kod97},
accounts also for the kinetic energy of the gas 
in the bubble. In this approach we develop further the observation 
made earlier regarding the homologous bubble shape \cite{Chu96},
but we allow for an adiabatic change of the bubble density function 
between the collapse and expansion phases. 
Our sole present objective is to establish that 
there is rapid transfer of  energy in sufficient quantity 
to the bubble, and that the bubble energy
quenching occurs equally rapidly on sub-nanosecond 
scale. In this way we show that the required energy focusing 
can be accomplished just in a short time interval 
without inception of shock waves in the
bubble, and that it arises from a combination of
highly non-linear bubble dynamics and rapid quenching by 
acoustical radiation of  bubble energy content \cite{Elz98}.

Following a general discussion of the shock wave mechanism in the 
SBSL phenomenon in section 2, we develop in section 3  
the variational hydrodynamic description of
the gas in the interior of the bubble. In section 4 we 
present numerical solutions and we study the  dynamics
of the energy flow to the gas bubble, and explore the dynamics
of matter and energy within the bubble. We summarize our 
results in section 5.

\subsubsection*{2. Shock waves and  light pulses in SBSL}
Perhaps the most popular proposed SBSL explanation up to this day involves the hypothesis that in order to generate 
the ultra-short 50ps light pulses seen in SBSL \cite{Bar97,Gom97}
one has to resort to extreme conditions that require shock wave 
formation in the bubble. Numerical 
hydrodynamic flow models incorporating inward shock formation 
\cite{WuR93,Mos94}, as well as possibly  outward motion shocks
\cite{Chu97} have been studied.
Despite considerable effort it is hard to prove convincingly that
shock waves are the power-source behind SBSL \cite{MCY97}. This is so, 
since there is no clear understanding how the shock motion energy 
is converted into  light pulses that depend very sensitively on
the atomic composition of the bubble. The only reason we see for shock 
mechanism popularity is that the extreme matter conditions associated
here with SBSL  naturally last for a relatively short time period,
well within the lifespan of the SBSL light pulse, which were originally 
reported to be less than 50ps long \cite{Bar97}. More recently, 
depending on the details of the SBSL conditions, pulse-length  
in excess of 100ps has been reported  \cite{Gom97}. 

Since we can obtain such time scale in section 4 without shocks,
one must ask  if the shock hypothesis should not be abandoned,
for there is a serious internal inconsistency: only perfectly
spherical inward shock could be reflected without destroying the
bubble. However, light emission asymmetry has now been observed 
 \cite{WPB96}: there is a class of SBSL events in which significant 
light ellipticity in the bubble is required, at the level of 20\%. It is
natural to associate this asymmetry with matter distribution asymmetry.
But then an inward shock wave could not bounce, the bubble matter 
would spray outwards, the bubble would burst. Even if a bubble were
to be rapidly (i.e. within a cycle of the driving harmonic 
pressure field) recreated, there  remains a contradiction with 
the well advertised magic of SBSL, that synchronous light 
pulses are emitted  over billions of shock cycles with timing
precision exceeding significantly that of the driving 
acoustical field generator. We recall that for this reason 
SBSL has even been proposed as 
a possible cheap and precise clock \cite{PutPub}. 

Should  we accept as argued that inbound shocks 
destroy the  bubble, they will also 
erase bubble `memory'. The enhancement of 
temporal precision of cycle definition would have 
to arise from the liquid alone, a highly improbable situation,
for which a mechanism is not in sight. An outward shock would not 
need to be reflected, but then the light pulse would be 
produced at the interface between gas and liquid, which is
also inconsistent with the experimental results \cite{WPB96}.
In our opinion the precision of light pulse periodicity 
can only arise from  nonlinear
dynamics within the gas bubble to which the external 
pressure field in the liquid is merely an  
energy source and hence it can be oscillating at a less well determined 
frequency. 

But what mechanism can produce light, if there is no 
shock? It is natural to expect that the light emitted is product 
of dynamical processes in the bubble.
Any microscopic light production mechanism
which does not generate strong spectral lines is suitable,  
including radiation   from  neutral atom
collisions with continuum emissions\cite{Fro94}. This 
collision induced emission (CIE) 
mechanism for generation of SBSL-light pulses appears 
most consistent with the reaction framework we are proposing here. 
CIE process relies on  relatively frequent and
sufficiently energetic atomic collisions 
in high density matter. To address 
qualitatively the viability of this mechanism, we establish below
in more detail the frequency of two-body atomic collisions
at relative atomic energy that would allow the light pulse formation.
While we do not pursue further more involved atomic physics aspects
of the CIE process, we 
note that the densities reached in the bubble at sonoluminescence 
implicate three and more body collisions.

We note that a 
bubble with radius similar to light wavelength is not
opaque to radiation, which thus is not emitted from a 
surface. Light emitted from the bubble center will not 
show features characteristic of being in equilibrium with
the  (thermal) source. Therefore the  conclusion  
that thermal radiation is 
excluded \cite{Hil98}, is not at all providing evidence for
shock-waves  as the mechanism of SBSL. 

Another non-shock process has been proposed as mechanism for SBSL,
involving strong, jet-like deformations invoked by the translational
motion of the bubble in the gradient of the acoustical field 
\cite{Pro97}. While jets 
indeed are likely to develop as the bubble begins to collapse,
in our opinion the long term bubble stability implies 
here that the 
increasing  internal bubble density and the increasing surface 
tension are capable to stabilize
the spherical shape of the bubble preceding the short final
 collapse phase, during which the SBSL light is emitted \cite{WBP97}. 
Were this not the case, the jet would have to excite SBSL in the 
final stages of bubble collapse with  a non-negligible 
chance for bubble destruction.

\subsubsection*{3. Variational bubble hydrodynamics}
Our theoretical  extension  of the standard RP approach 
concerns the dynamics of an `adiabatically'
homologous gas flow. This means that the shape and velocity 
of the interior matter are governed by the rapidly evolving  
radius parameter $R(t)$ and surface velocity $\dot R(t)$,
allowing for slower change in the homologous shape of matter 
characterized by a collective shape parameter $a(t)$. 
In order to exploit this to the fullest, we implement the 
dynamics within the variational hydrodynamical 
method \cite{Kod97}, which has 
been  discovered nearly 40 years ago (see Ref. \cite{YM79}, 
section 13, and  references therein). Using the
variational approach we sidestep the need for a full
hydrodynamic solution of the gas motion. This simplified 
approach is suitable only for exactly spherical, 
(quasi) one-dimensional motion, and  corresponds to a variational
solution of the hydrodynamic problem with irrotational flow.
 
We consider a gas bubble of radius $R(t)$, shape $g(x,a(t))$
mass $M_{\rm G}=\frac43\pi R_0^3\rho_0$ (expressed in terms 
of the initial radius $R_0$ and density $\rho_0$) surrounded by an incompressible liquid of density $\rho_{\rm L}$. 
In terms of the  $R$-scaled radial variable, the  mass-conservation 
constraint,  $\dot M_{\rm G}=0,\,$ implies:
\begin{eqnarray}
\label{scale}
\rho_{\rm G} =\frac {M_{\rm G}}{4\pi R(t)^3}g(x,a(t))\,,&&\quad
x=\frac r{R(t)}<1\,,\\
\int_0^1 x^2g\,dx=1\,,\qquad &&
\int_0^1 x^2\frac{\partial g}{\partial a}\,dx=0\,.
\label{Normg}
\end{eqnarray}
$R(t)$ and the collective variable $a(t)$ are the dynamical 
functions we will
consider. $a(t)$  comprises two important physical phenomena:
\begin{enumerate}
\item the slow underlying variation (`drift')
 describes how the shape of the 
mass density in the bubble changes between the contraction and 
expansion phases of the bubble motion -- solving at fixed 
$a=$Const. We are implying that the shape shift  is adiabatic
compared to the rapid matter flow, expressed by changes in $R$; 
\item small, but rapid oscillation of $a(t)$ may be
superposed with this slow drift; they correspond to 
the acoustical cross-talk oscillations between
the surface and the interior of the bubble with
frequency $\omega\simeq c_s/R\le10^9$s$^{-1}$, 
where $c_s$ is the sound velocity in the gas.
\end{enumerate}
The introduction of
$a(t)$ rather than time $t$ in definition of the 
shape function $g$ 
in Eq.\,(\ref{scale}) serves to separate these two 
types of non-homologous dynamics of the bubble. 

In order to determine the  dynamics of the gas bubble,
we will seek an extremum of the action for the combined 
gas-liquid system using as dynamical variables 
$R(t),\,\dot{R}(t)\, \dot a(t),\, a(t) $\,. The matter
shape function $g(x,a)$ is imposed by the external constraints.
The  kinetic energy of the gas can be easily expressed as a 
function of $g$: solving the continuity equation we obtain
the local flow velocity in the gas bubble:
\begin{equation}\label{flow}
v(r)=x\dot R-R\dot a\delta\,,\qquad
\delta =\frac 1{x^2g(x;a)}\int_0^x\! dx'
    x'^2\frac{\partial g}{\partial a}\ 
{\longrightarrow 0}_{\!_{\hspace*{-0.9cm} x\to 1}}\,
\hspace*{1cm}.
\end{equation}
After some integrations by parts of the time integral (action) 
the contribution in action is found \cite{Kod97}:
\begin{eqnarray}\label{gasenergy2}
K_{\rm G} &=&\frac{M_{\rm G}}2\left\{ 
- I_1R\ddot{R}  +I_3\dot{a}^2R^2\right\}\,,
\end{eqnarray}
with:
\begin{eqnarray}\label{I123}
I_1=\int_0^1dx\,x^4g\,,\qquad
I_3 =\int_0^1dx\,x^2g\,\delta^2 \,. 
\end{eqnarray}
When $\delta$ is neglected ($\frac{\partial g}{\partial a}\to 0$)
one is in the exact homologous limit.

Eq.\,(\ref{gasenergy2})  complements the kinetic energy of the 
incompressible liquid of density $\rho_{{\rm L}}$. Solving
the continuity equation we obtain the local flow velocity and thus
also the kinetic energy in the liquid:
\begin{equation}\label{Kliq}
v_{\rm L}=\dot R \frac{R^2}{r^2}\,,\quad r>R\,;\qquad
K_{{\rm L}}=4\pi\rho_{{\rm L}}R^3\frac{\dot R^2}{2}\,.
\end{equation}
The  internal (potential) energy of the gas is:
\begin{eqnarray}\label{Wgas}
W_{\rm G}&=&W_{\rm G}[R,g(x;a)]=\int \varepsilon (r,t)\;d^3r 
      =M_{\rm G}\int_0^1\!dxx^2g{\cal E}\,,  
\end{eqnarray}
where $\varepsilon $ is the energy density and 
$${\cal E}\equiv E/N=\varepsilon /\rho  $$ 
is the specific energy of the gas. The total potential
energy $W$ of the bubble (normalized to zero for a bubble
at rest under ambient conditions (pressure $P=P_0$ and temperature
$T=T_0$) is: 
\begin{equation}\label{poten}
W=W_{\rm G}+4\pi \sigma (R^2-R_0^2)-P_{{\rm L}}(V_{{\rm 0}}-V_{{\rm G}})\,. 
\end{equation}
We note that the second 
term in Eq.\,(\ref{poten}) is the energy from the fluid-gas
interface, and $P_{{\rm L}}=P_0+P_{\rm a}(t)$ 
is the pressure in the liquid
arising from the atmospheric pressure and any further time
dependent externally applied (acoustical) pressure. 
Setting $I_1=I_3=0=K_{\rm G}$ i.e.  ignoring the interior dynamics of 
the bubble, and varying $K_L-W$
with respect to $R(t)$ one obtains the (improved) 
RP-model equations \cite{Lof93}\,.
 
We first determine the dynamical equations which
characterize the
drift in $a(t)$. The tool of convenience here is the 
Hamiltonian for the motion of this dynamical variable. 
The generalized momentum following from 
Eq.\,(\ref{gasenergy2}) is:
\begin{equation}\label{momentuma}
\Pi_a=\frac{dK_{\rm G}}{d\dot a}=\dot a I_3(a)R^2M_{\rm G}\,,
\end{equation}
and the Hamiltonian for the gas is then
\begin{equation}\label{hamila}
{\cal H}_{\rm G}(a,\Pi_a)=\frac12 \frac{\Pi_a^2}{\cal M}_a
+{\cal V}_a\,.
\end{equation}
Where the (collective) mass  ${\cal M}_a$
and potential ${\cal V}_a$ are:
\begin{equation}\label{MV}
{\cal M}_a\equiv {I_3(a)R^2M_{\rm G}}\,,\quad
{\cal V}_a\equiv 
\frac{M_{\rm G}}{2} I_1(a) R\ddot R + 
M_{\rm G}\int_0^1\!dx  x^2g(a){\cal E}[\rho(a)]\,.
\end{equation}
where $I_1,\,I_3$ are given in Eq.\,(\ref{I123}). 
The rapid oscillations inherent in 
the harmonic Hamiltonian presented are centered around the
minimum of ${\cal V}_a$, which drifts as the values of
$R\,,\ddot R$ change. At the same time there is a drift in the 
mass ${\cal M}_a$ which impacts the frequency of the rapid
oscillations, but not the location of the minimum of ${\cal V}_a$.
thus if we study solely the drift in $a$-coordinate, we are
effectively neglecting the deviations from the homologous 
shape inherent in the $\delta$-dependence of $I_3$.

As we search for the minimum of the effective 
potential in order to derive equations governing this 
motion we will need to invoke a few simple thermodynamic
properties. From the first law of thermodynamics we recall the
relation 
of pressure $P$ to the internal specific energy ${\cal E}$:
\begin{eqnarray}\label{Prho}
P=-\left(\frac{\partial E}{\partial V}\right)_{|S,N}
=\rho^2\left(\frac{\partial{\cal E}}{\partial\rho}\right)_{|S}\,.
\end{eqnarray}
The heat function (enthalpy) $h$ is introduced in view of the
relation:
\begin{eqnarray}
  \frac{d}{d\rho}({\cal E}+\frac{P}{\rho})\left.\right|_{S}=
    \frac1\rho\frac{dP}{d\rho}\left.\right|_{S}\,
         \quad \longrightarrow \qquad
h\equiv {\cal E+}\frac P\rho = 
     \int_{S=Const}\frac{dP}\rho\,.\label{h}
\end{eqnarray} 
The minimum of ${\cal V}_{\rm a}$ determines the drifting value of 
$a$\,, which is obtained by solving:
\begin{equation}\label{drifta}
\frac{1}{M_{\rm G}}\frac{d{\cal V}_{\rm a}}{da}=
\int_0^1\!dx\, x^2\,\frac{\partial g}{\partial a}
\left\{h[\rho ]-h_c+\frac12 x^2 R\ddot R\right\}=0\,.
\end{equation}
Here we have introduced a constant $h_c$, 
which integrates to zero
by the second expression in Eq.\,(\ref{Normg}), and which assures that  
 the solution of Eq.\,(\ref{drifta}) is found where the curly bracket in 
Eq.\,(\ref{drifta}) vanishes. Considering the limit
 $x^2\to 0$ we see that $h_c=h(x=0)$.
Differentiating with respect to $x^2$ the argument of the 
curly bracket we obtain, under
adiabatic constraints and using the 
definition of $h$, see Eq.\,(\ref{h}):
\begin{equation}
-\frac12 R\ddot R 
=\frac{d}{d x^2}\left\{h[\rho ]-h_c\right\}
=\frac{\partial\rho}{\partial x^2}
\frac{\partial P}{\partial \rho}\frac{\partial h}{\partial P}
\,,
\end{equation}
and thus:
\begin{equation}\label{intshape}
\frac{\partial}{\partial x^2}\ln g(x;a)
=-{{ R\ddot R} \over {2c_s^2(\rho(g))}}\,,\qquad 
c_s^2(\rho)=\frac{\partial P}{\partial \rho}\,,
\end{equation}
where $c_s$ is the local sound velocity in the gas. Here the 
factor $R\ddot R$ is to be seen as a given parameter; in reality
it is determined by the usual RP-dynamics, which does not depend 
much  on  the inclusion in the dynamics of the  bubble interior
kinetic energy of the. Thus 
Eq.\,(\ref{intshape}) can be solved independently 
of the RP-dynamics, yielding the interior shape of the bubble at 
each phase of the dynamical evolution of bubble radius.
 We note that  the sign of the RP-force factor 
$R\ddot R$ in Eq.\,(\ref{intshape})
 determines the gradient of the density shape function: 
when the bubble is 
collapsing, the surface  density is higher than
the interior density, while when it is expanding the 
opposite is the case, and a flat distribution is found 
in the force free  instants of the bubble motion, when 
$\ddot R=0$. We have explored elsewhere other theoretical features 
not discussed in depth here, such as the small fluctuations 
in $a(t)$ \cite{Kod97}, the back-reaction of the bubble 
dynamics into the RP-equations \cite{Kod97}, 
non-perturbative treatment of the sound emission\cite{Elz98}.
Also, we have obtained  Eq.\,(\ref{intshape}) for $g(x,a)$ 
directly from the Euler equation in the limit $\delta\to 0$\,.
 
We note that only for a particular choice of the
initial condition g(x=0) 
can  we obtain a properly normalized solution of the nonlinear
differential equation, Eq.\,(\ref{intshape}). 
However, this we can do easily for any EOS subject 
to an adiabatic constraint,
and specifically also for the van\,der\,Waals system we consider here, 
with finite size correction to the volume in the polytropic
 equation of state:
\begin{equation}
P[\rho ]=P_0\left( \frac{\rho _m}{\rho _0}-1\right)^\gamma 
     \left( \frac \rho{\rho_m-\rho }\right)^\gamma ,
\label{P-VW}\quad 
h[\rho]=\frac 1{\gamma -1}\frac P\rho \left[ \gamma 
    -\frac \rho {\rho _m}\right]\,, 
\quad {\cal E}[\rho]=\frac{P}{\gamma-1}
\left(\frac{1}{\rho}-\frac{1}{\rho_m}\right)\,,
\end{equation}
with $\rho _m\equiv M_{\rm G}/a^3=3M_{\rm G}/4\pi r_m^3$ and $r_m$ is the 
van\,der\,Waals excluded radius.

\subsubsection*{4. Physical properties of sonoluminescing bubbles}
The motion of the bubble is primarily damped by 
sound radiation in the crash phase. We have determined 
that in rough detail a full treatment of the sound radiation 
is equivalent to first order in $\dot R/c$ description 
when  adjustment (i.e. fit) of the damping constants
is allowed for \cite{Elz98}. We thus follow
the approach of L\"ofstedt \cite{Lof93} and account for 
the sound damping influence on dynamics of $R(t)$  by including the 
appropriate  first order in $\dot R/c$ terms   in our
extended Rayleigh-Plesset equation. We also incorporate the damping  
due to liquid viscosity. We do not study  the effect of 
compressibility of the fluid surrounding the gas bubble. 
We  use the parameters adjusted in Ref.\,\cite{Lof93} to fit the dynamics of the bubble radius. 

The standard example we consider here is an argon
gas bubble comprising $N_{\rm G}=10^{10}$ atoms immersed
in water. The parameters of the liquid are: density
$\rho_{\rm L}=1$\,g/cc\,, surface tension
$\sigma=0.03\,\rm{kg/s}^2$\,,
sound velocity $c_{H_2O}=1450$\,m/s\,, and viscosity 
$\eta=3\cdot 
10^{-3}$\,kg/(ms)\,. The equilibrium radius of a bubble
under ambient pressure $P_0=1$\,atm 
at ambient temperature $T_0=277K$ 
obtained from the van\,der\,Waals equation state 
is $R_0=4.39\,\mu$m and
the excluded radius parameter is 
$a=0.503\,\mu$m\,. The acoustical driving pressure is 
taken to be $P_{\rm a}(t)=P_{\rm a}\sin (\omega t)$, 
with $P_{\rm a}=1.4\,$atm and  
$\omega /2\pi =25$\,kHz\,.  

Our numerical procedure is straightforward: we determine the
 solutions of the RP-equations,
and use these to solve Eq.\,(\ref{intshape}) at a given instant in time. The solutions turn out
to be very smooth in time and space 
and we can interpolate quite easily
the solutions with large time steps. 
Subsequently, we can study any 
property of the bubble.

When we  evaluate, as a function of time, the energy content in the
bubble, we find that there are three time scales
of interest in the gas bubble dynamics:
the overall energy of the liquid and the bubble 
varies on scale of driving frequency, within the collapse 
window of a few microseconds, as can be 
seen in  Fig.\,1a. 
Note that we present the energy of the liquid per 
atom in the gas bubble.  The solid line is the total energy, the 
dotted line is the kinetic energy of flow, the dashed line is the 
internal (potential) energy arising from the work done on the liquid. 
Both components contribute significantly to the overall energy of the 
liquid. We see
that the potential energy dips during the slow expansion phase of
the bubble motion, and then smoothly rises to a maximum value at
which point it declines at a slightly faster rate. During the
collapse the kinetic energy
of the fluid rises to ${\cal O}$(60~eV) per atom in the bubble gas,
which energy is rapidly lost to elastic and inelastic 
processes when  the bubble hits the van der Waals hard core.
The dynamical time scale of the fluid, defined as the time difference
in minimum-maximum-minimum energy evolution, is ${\cal O}(10\,\mu$s), 
corresponding to a quarter of the acoustical driving cycle
of 40$\mu$s. The interior gas energy varies on the scale of
nanosecond, as can be seen in  Fig.\,1b. 
In this example about 7\% of the energy is transferred 
from the liquid to the gas. Finally, in Fig.\,1c 
we show a time-resolved view of the energy contents of
the interior of the bubble.
We see that most of the energy rise occurs within a 100 picosecond 
time window. The energy of the bubble disappears rather slowly;  
there is no viscous damping included
in the dynamics of the gas.
 
We now turn our attention to estimate the intensity of the
sonoluminescence arising from the atomic collision processes; 
for this we obtain the collision frequency 
of all atoms in the bubble: 
\begin{equation}\label{Bcol}
f_N=N\sigma \frac{\rho}{M_{\rm G}} v_{\rm rel}\,.
\end{equation}
The effective collision cross section is assumed to be geometric and
thus  $\sigma \simeq (4\pi/3a^3/N)^{2/3}$\,. 
A measure of relative velocity is obtained from 
the total energy per atom $v_{\rm rel}\simeq \sqrt{2E/m}$.
We can obtain these results for a number of different conditions
of driving pressure, ambient pressure, frequency, etc. In Fig.\,2
we present the results for $P_a=1.4,\,1.3,\,1.2$\,atm, selecting an
arbitrary time axis origin so that these results can be shown together.
The number of collisions shown is in $10^{10}/$ps, thus during the light
flash duration of 50\,ps  we expect about $10^{12}$--$10^{13}$
atom-atom collisions to occur in the bubble. We recall that
experimentally observed is the emission of up to 
million photons in the spectral range 
of visible light during the same period of time 
(see e.g. legend of figure 1 in Ref.\cite{Bar97}.) 
We also see in Fig.\,2 
that the number of collisions is 
changing smoothly, and 
in a manner corresponding to the emission intensity observed in
the experiment \cite{Bar97}. The vertical lines indicate when in time
the average energy per atom exceeds 1 eV.  The realm of the 
frequent collisions at 1 eV and above is well within the 500ps time
interval of the most extreme density,
as is experimentally reported for SBSL \cite{Bar97}. 
In our opinion these results strongly support 
the hypothesis that atomic collisions 
are the source of the  short SBSL light 
flashes \cite{Fro94}, since the number of atomic collisions
capable to generate visible photons exceed by the comfortable
margin of $10^6$--$10^7$ the photon yield. 

While the onset of SBSL is a gradual process, depending on
the magnitude of the driving acoustical pressure field, and could be
explained as the onset of frequent high energy atomic collisions, 
the disappearance of SBSL when $P_a> 1.4 $\,atm is quite sudden.
We  considered the local sound velocity $c_s(x)$ in the bubble and 
compare it to the local flow velocity $v$, Eq.\,(\ref{flow}), and 
have determined  that this condition corresponds to the interior
flow velocity in the gas bubble slightly exceeding, near to the 
surface, the local sound velocity.
The flow velocity in the bubble, in units of local sound 
velocity, is shown for several values 
of driving pressure $P_a=1.4,\,1.3,\,1.2$\,atm
in Fig.\,3 
As $P_a\to 1.4$ atm, we see in Fig.\,3 
that the local flows in the bubble 
are bound to occur with supersonic velocity. For the development of 
shock waves just this supersonic condition is required. Since 
in the experiment $P_a\simeq 1.4$ atm is the upper limit 
for the SBSL  process, we are led to the hypothesis that it is 
the onset of shock  instability in the bubble, which is responsible 
for the upper limit  in $P_a$ for which  SBSL can occur.

\subsubsection*{5. Discussion and conclusions}
In summary, we have developed a semi-analytical dynamical model 
to describe the properties of the interior of the cavitating 
gas bubble. In our approach we have simplified the complexity of 
the hydrodynamic gas flow problem by exploiting  in a variational 
approach the nearly homologous  bubble dynamics. Considering
that SBSL occurs in an extremely small 
domain  of a large parameter space, our approach is in 
some aspects more capable to gain physical insights, than 
is full hydrodynamics. Several approximations were made:
we did not address quantitatively in our 
numerical work the sound
fluctuations inside the bubble during its motion.  We did not
consider here the back-reaction from the 
dynamics of the gas bubble to the dynamics of the liquid which, 
while important in principle, does not change the general 
dynamic behavior of the  bubble \cite{Kod97}. Neither did
we include the compressibility in
 the description of the surrounding fluid 
motion, but our state of the art implementation of the dynamics of 
the bubble includes in a way proposed in Ref. \cite{Lof93}
the dissipative effects such as  
viscosity of the fluid, and most importantly, 
acoustic radiation damping \cite{Elz98,Lof93}.
We checked and agree with Ref. \cite{Gom97} that allowing for 
compressibility of the liquid in the extreme conditions
of SBSL further sharpens the characteristic time
scale involved, as the denser liquid can more rapidly 
acoustically  dissipate the air bubble energy. We find  
that even without this effect  there is a natural 100-picosecond
time scale in argon cavitation in water, arising from the
transfer of kinetic flow energy
from the liquid to the gas bubble in its expansion phase, 
followed by inertial collapse of the energized bubble. 

We were able to 
derive the density shapes of the bubble and to evaluate
the energy content of the bubble as function of time. 
We have shown that the bubble energy has a characteristic
relaxation time  ${\cal O}($1\,ns), while  the scale 
${\cal O}$(100\,ps) describes  the 
rise in energy of the bubble. 
We have shown in quantitative fashion that it is 
likely that atomic collisions are the source of  the
light flashes seen in SBSL. 
We refrain here from the evaluation of the CIE rates 
\cite{Fro94}, pending further clarification of 
microscopic mechanisms (specifically   the role of noble 
gases, three body collisions), and also since this task 
transcends the scope of this research 
project. Thus while this is not the time and
place to explore in  detail the light spectrum that can be produced,
we note that during the short time interval allowing 
deep atomic collisions, CIE will produce light pulses of similar 
temporal length  for different wavelengths in the visible range. 

The dynamical properties of the RP-model lead us 
to believe that it is the (inward) shock instability that poses an
upper-applied-acoustical-pressure-limit 
on the SBSL phenomenon. Specifically,  when the 
strength of the driving acoustical pressure reaches the value
at which experimentally SBSL phenomenon suddenly disappears \cite{Bar97},
our calculations show that the internal
bubble dynamics must evolve faster than local sound velocity.
This should lead  to flow instabilities, which in all likelihood 
can destruct the bubble. Our results imply
that the complex, near-shock-nonlinear dynamics of the 
bubble interior  combined with light emissions in 
atomic collision processes  
can provide a suitable framework to explain the
SBSL phenomenon.

\subsection*{Acknowledgment}
Work partially supported by: DOE, grant DE-FG03-95ER40937; by NSF,
grant  INT-9602920; and by PRONEX-41.96.0886.00 -- Brazil.

\newpage
\vfill
\begin{figure}[tbh]
\vspace*{1.0cm} 
\centerline{\hspace*{-1.6cm}
\psfig{width=27cm,figure=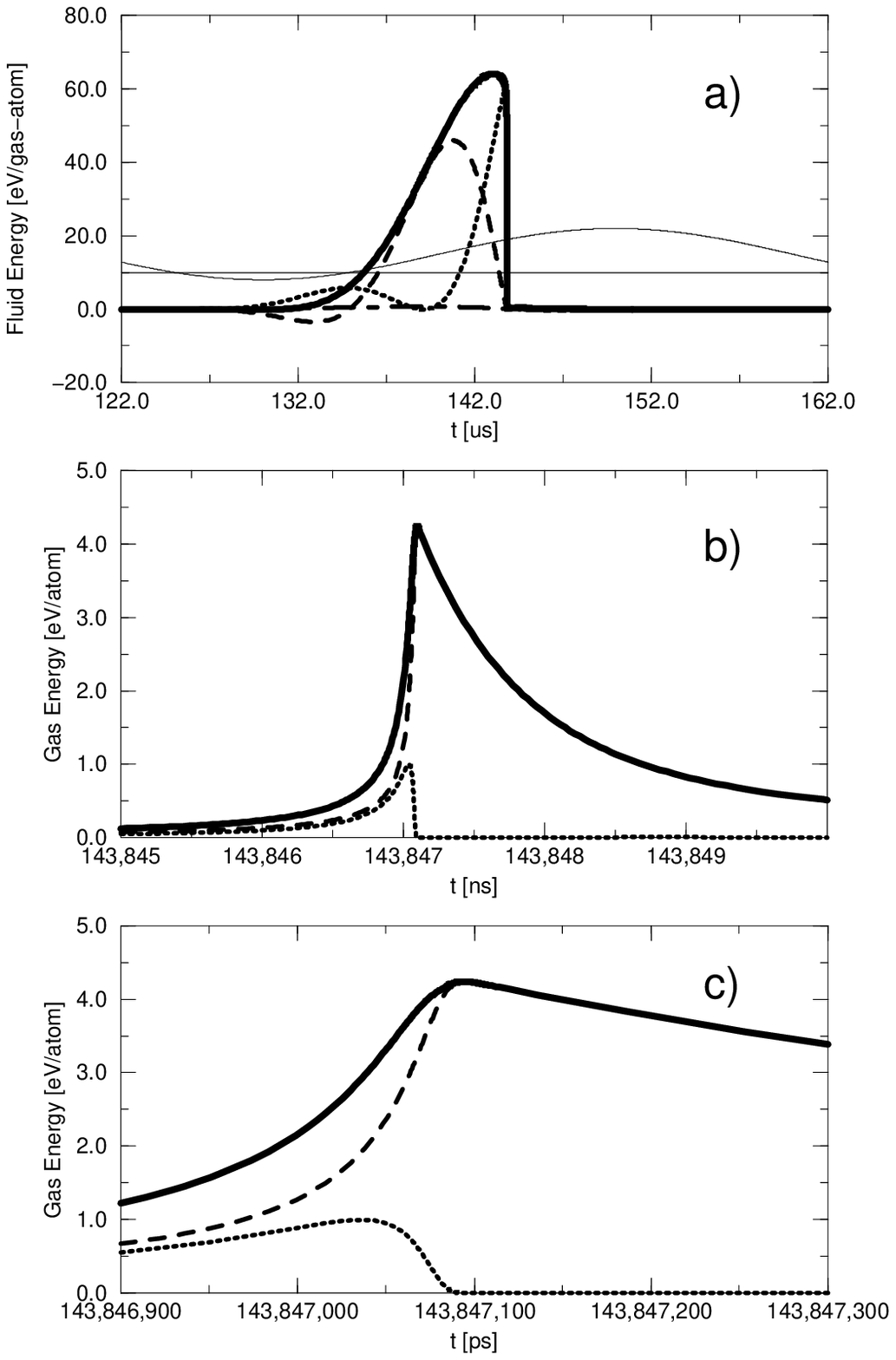}
}
\vspace*{-1.6cm}
\caption{Argon bubble with \protect{$10^{10}$} atoms in water.
Evolution of gas bubble energy: dotted line is the kinetic energy, 
dashed line is the
potential energy, and solid line is the total energy. 
The thin solid line over arbitrary
scale shows the sum of ambient  pressure, 1\,atm, and driving acoustical
 pressure (\protect{$\omega=25$}\,kHz), amplitude at bubble surface  
1.4 atm:
a) The energy in the fluid in eV per atom in the bubble;
b) The energy in the bubble over nanosecond time scale;
c) The fine resolution in time of the bubble energy during most extreme 
   collapse.
\protect\label{Bubble}}
\end{figure}
\eject
\begin{figure}[tbh]
\centerline{\hspace*{-1cm}
\psfig{width=17cm,figure=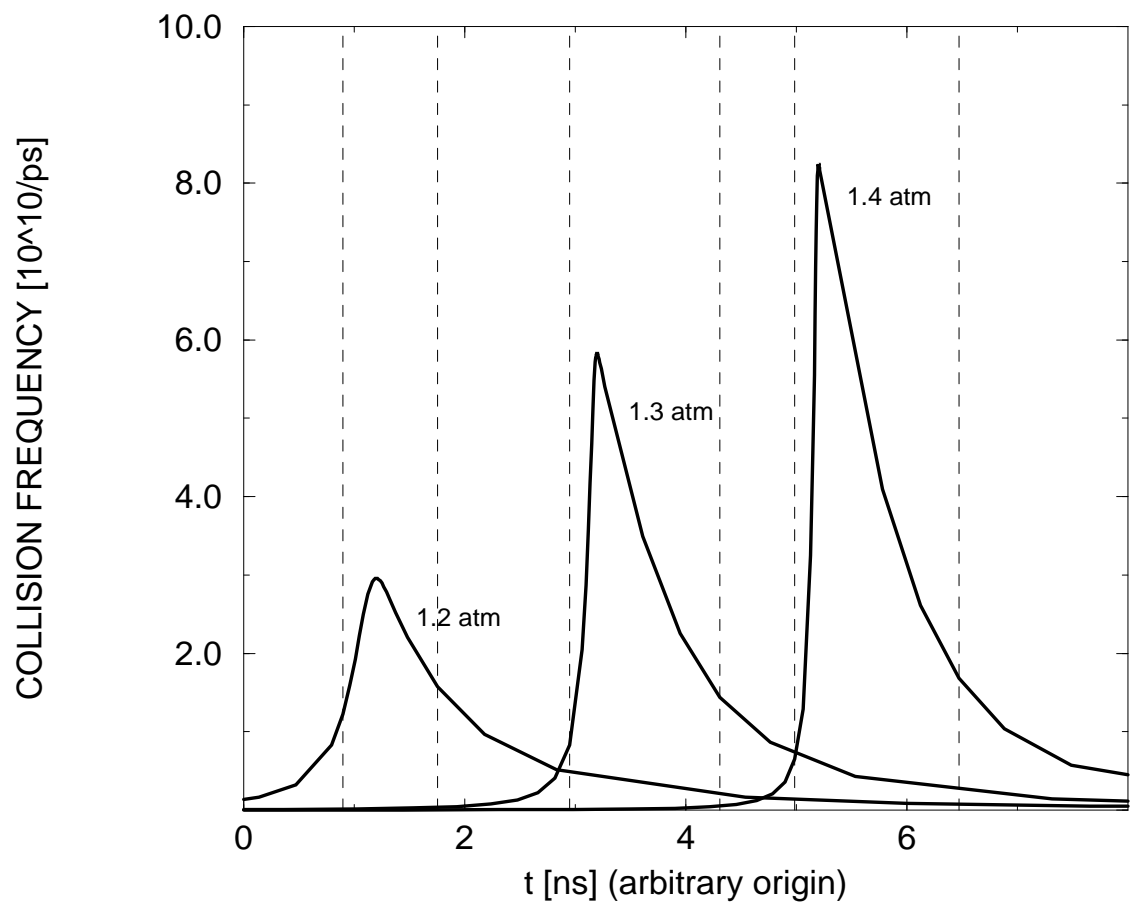}
}
\vspace*{-1cm}
\caption{Number of atom-atom collisions (times \protect{$10^{10}$})
per picosecond in the bubble as function 
of time for three values of driving acoustical amplitude, 
\protect{$P_a=1.2,\,1.3,\,1.4$}\,atm.  Dashed vertical lines delimit 
range for 1 eV energy per atom in the bubble.
\label{Bfreq}
}
\end{figure}
\eject
\begin{figure}[tbh]
\centerline{\hspace*{-1.5cm}
\psfig{width=16cm,figure=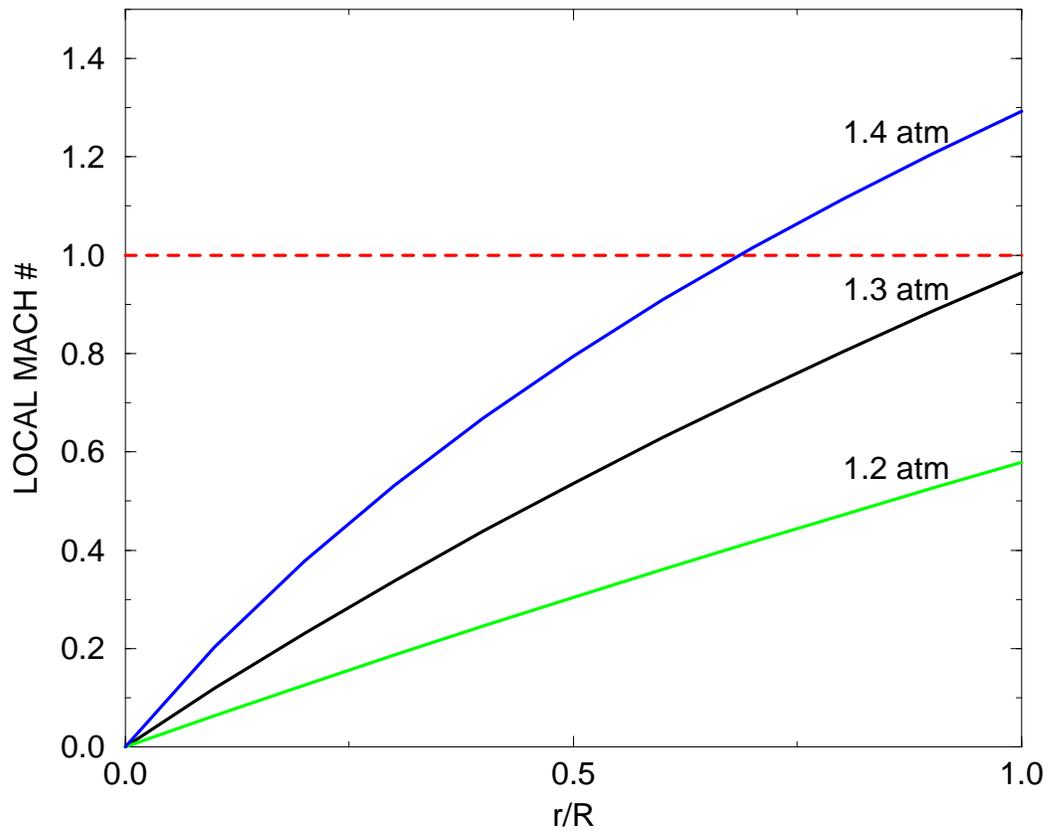}
}
\vspace*{-1cm}
\caption{Local flow velocity in units of the local sound velocity 
(termed Mach \protect\#), as function of the scaled radial coordinate
 of the bubble, \protect{$r/R$}. Dashed line: Mach \protect\#=1.0\,. 
Results for  driving pressure \protect{$P_a=1.4\,,\ 1.3\,,\ 1.2$}\,atm 
are shown (top to bottom). Event time chosen to maximize the local Mach 
\protect{\#} at the bubble surface.
\label{BBspeed}
}
\end{figure}

\end{document}